\documentclass[paper]{geophysics}
\usepackage{mathrsfs,amsmath, multirow, xspace}
\usepackage{hyperref}
\usepackage{makecell}
\begin{document}

\title{Open source software for simulations and inversions of airborne electromagnetic data}

\renewcommand{\thefootnote}{\fnsymbol{footnote}}

\ms{...} % manuscript number

\address{
\footnotemark[1]Geophyscial Inversion Facility, \\
University of British Columbia, Canada\\
BC, Vancouver}

\author{Lindsey J. Heagy\footnotemark[1],\footnotemark[2] Seogi Kang\footnotemark[2], Rowan Cockett\footnotemark[2] and Douglas W. Oldenburg\footnotemark[2]\\[16pt]
% affiliations
{\normalfont \small
\footnotemark[1]
Corresponding author: lheagy@eos.ubc.ca \\
\footnotemark[2]
Geophysical Inversion Facility, University of British Columbia
}\\[16pt]
% keywords
{
\normalfont \small
}
}

\footer{Open source software for AEM}
\lefthead{Heagy et al., 2018}
\righthead{Open source software for AEM}

\maketitle

\begin{abstract}
Inversions of airborne EM data are often an iterative process, not only requiring that the researcher be able to explore the impact of changing components such as the choice of regularization functional or model parameterization, but also often requiring that forward simulations be run and fields and fluxes visualized in order to build an understanding of the physical processes governing what we observe in the data. In the hope of facilitating this exploration and promoting reproducibility of geophysical simulations and inversions, we have developed the open source software package, SimPEG. The software has been designed to be modular and extensible with the goal of allowing researchers to interrogate all of the components and to facilitate the exploration of new inversion strategies. We present an overview of the software in its application to airborne EM and demonstrate its use for visualizing fields and fluxes in a forward simulation as well as its flexibility in formulating and solving the inverse problem. We invert a line of airborne TDEM data over a conductive vertical plate using a 1D voxel-inversion, a 2D voxel inversion and a parametric inversion, where all of the forward modelling is done on a 3D grid. The results in this paper can be reproduced  by using the provided Jupyter notebooks. The Python software can also be modified to allow users to experiment with parameters and explore the physics of the electromagnetics and intricacies of inversion.
\end{abstract}

\section{INTRODUCTION $\&$ MOTIVATION}
Even for a simple model of a conductive vertical plate in a resistive half-space, the contributions of the diffusing ``smoke ring'' currents which intersect the plate, and vortex currents induced by the time-varying magnetic flux through the plate to the observed airborne EM signal can be challenging to unravel. These challenges are exacerbated if the background has significant structure or if other physical properties such as magnetic permeability or chargeability are significant. Dissecting and understanding the physical phenomena that produce the data which we observe in an airborne EM survey requires that we have the ability to simulate Maxwell's equations and visualize currents and magnetic fields through time or over a range of frequencies for a variety of models, which might be 1D, 2D, or 3D in space.

Beyond building a fundamental understanding of the physics governing an airborne EM response, extracting meaningful geologic information from airborne EM data requires that we have the ability to invert those data. EM inversions are nonlinear and thus the choice of regularization and selection of tuning parameters can have a significant impact on the recovered model. To progress research into extracting more geologic or hydrogeologic information from airborne in an inversion, researchers require the flexibility to experiment with aspects of the inverse problem including the regularization and tuning parameters, but also aspects such as the dimensionality of the numerical simulation and definition of the inversion model (e.g. voxel or a parametric model). Looking forward, we also want the ability to interface to other geophysical methods through joint or cooperative inversions.

The classic model of black-box, proprietary software does not promote the collaboration or transparency necessary to achieve these goals. In the hope of facilitating collaboration and easing the overhead for researchers in going from a new idea to an implementation of that idea, we have been developing an open source framework and software implementation for Simulation and Parameter Estimation in Geophysics, SimPEG \citep{cockett2015}. SimPEG includes finite volume simulations and inversion routines for a variety of geophysical applications including Potential Fields, Vadose Zone flow, DC Resistivity, and Electromagnetics. Simulations may be performed on several different mesh types, including cylindrically symmetric meshes, 3D tensor meshes and OcTree meshes. The fields and fluxes computed everywhere in the simulation domain are readily accessible so that they can be easily visualized and explored, particularly when used in the interactive Jupyter computing environment \citep{Perez2015}. Such simulations and visualizations have proved valuable in the context of geoscience education \citep{Oldenburg2017} and can be a useful tool for understanding the physical processes that contribute to the data we observe.

\cite{heagy2017} provide a complete overview of the structure and implementation of electromagnetic simulations and inversion within SimPEG. In brief, SimPEG includes staggered-grid, finite volume solutions to the quasi-static Maxwell's equations in both the time domain and the frequency domain on cylindrically symmetric, 3D tensor meshes, and OcTree meshes. Variable electric conductivity and magnetic permeability may be considered. \cite{kang2016} have recently extended the implementation to account for chargeable material.

We take a deterministic, gradient-based approach to the inverse problem and consider a Tikhonov style inversion, where the inversion minimizes an objective function consisting of a data misfit and a regularization functional. The inversion model is decoupled from the forward simulation mesh, allowing different model parameterizations to be employed. For example, a 2D inversion can be performed for a single line of airborne EM data while the forward simulation is conducted on a 3D mesh, or, similarly, a parametric model (e.g. a plate in a half-space earth) may be considered. SimPEG has been built in a flexible, modular manner, to enable researchers to experiment with components such as which norm is employed in the regularization or which mesh is used in the forward simulation.

SimPEG is implemented in Python and is licensed under the permissive MIT license which allows commercial and academic use and adaptation of the software (\href{https://simpeg.xyz}{https://simpeg.xyz}); we hope this maximizes the utility of the software itself and creates opportunities for new contributions.

In what follows, we will demonstrate the use of SimPEG for simulations and inversions of airborne EM data through three examples. In the first example, we will consider an airborne TDEM experiment over a vertical conductive plate and examine the behaviour of the channeled smoke ring currents and induced vortex currents and discuss how each contribute to the measured data. From there, we will demonstrate a 1D inversion of TDEM data collected over the plate. Our final example will consist of two 2D inversions where the forward simulations are carried out in 3D; the first inversion is for a voxel-based model and the second is for a parametric model. We will conclude with a discussion on extensions of this example, including our vision of how to move forward as we look to address large 3D airborne EM inversions using the SimPEG framework.

In the examples that follow, we will use a model of a conductive, vertical plate in a resistive half-space. This is a problem of relevance in mineral exploration, brings to light a number of fundamental aspects of electromagnetics, and is a challenging inverse problem, particularly as 1D inversions cannot explain the 3D nature of the model. We will look at the behaviour of currents through time and demonstrate the multi-dimensional nature of the EM responses and then conduct 1D and 2D inversions of synthetic airborne TDEM data over the plate. The examples have purposefully been kept computationally lightweight so that they can be run in a moderate amount of time on a modern laptop. The source code for each of these examples is available as a Jupyter notebook at
\href{https://github.com/simpeg-research/heagy-2018-AEM}{https://github.com/simpeg-research/heagy-2018-AEM} \citep{Heagy2018}.

\subsection{Airborne TDEM response over a conductive plate}
\begin{figure}[!htb]
  \centering
  \includegraphics[width=0.85\textwidth]{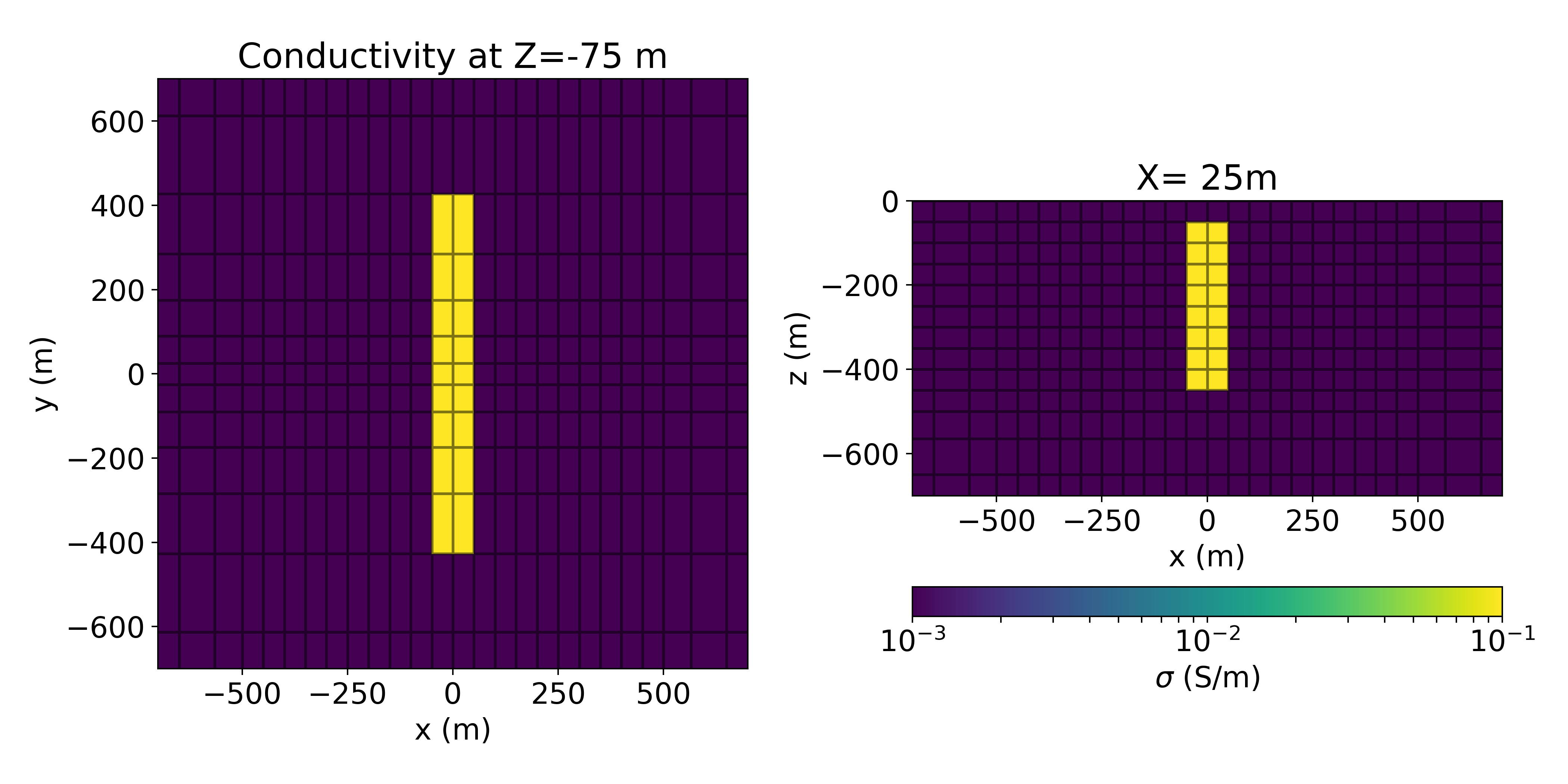}
  \caption{Depth slice (left) and cross section (right) through the model of a conductive plate (0.1 S/m) in a resistive half-space ($10^{-3}$ S/m).}
  \label{fig:model}
\end{figure}

The model and simulation mesh we consider is shown in Figure \ref{fig:model}. A 100 m thick conductive plate (0.1 S/m) is embedded 50 m below the surface. The plate is 400 m tall and extends 850 m along the x-axis. The background conductivity is $10^{-3}$ S/m. One line of airborne TDEM data is computed along y=0m; the source waveform is a step-off waveform, and vertical $db/dt$ data are collected at 21 time-channels between 0.05 and 2.5ms. The simulation is performed in 3D on a tensor mesh.

\begin{figure}[!htb]
  \centering
  \includegraphics[width=0.7\textwidth]{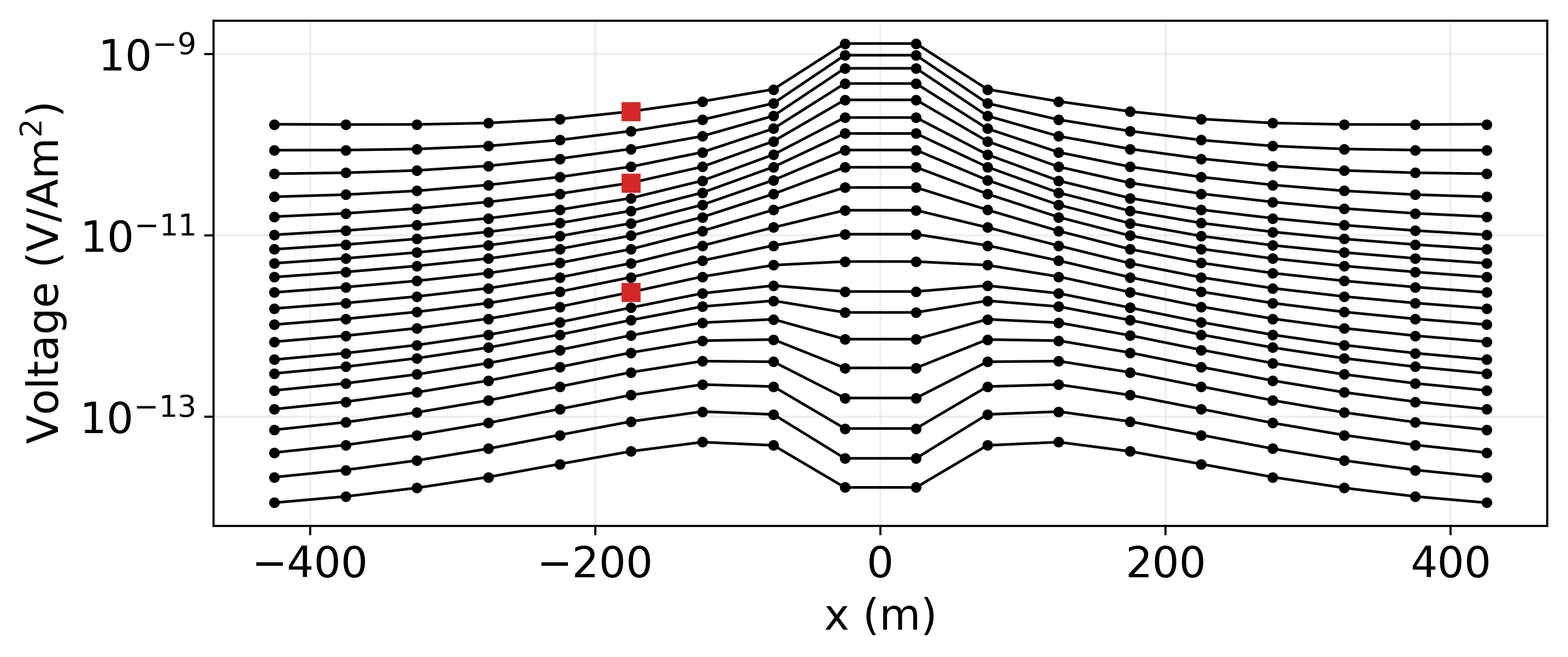}
  \caption{Vertical $db/dt$ data over the conductive plate. The red dots correspond to the times and source locations shown in Figure \ref{fig:currents}.}
  \label{fig:data}
\end{figure}

\begin{figure}[!htb]
  \centering
  \includegraphics[width=0.85\textwidth]{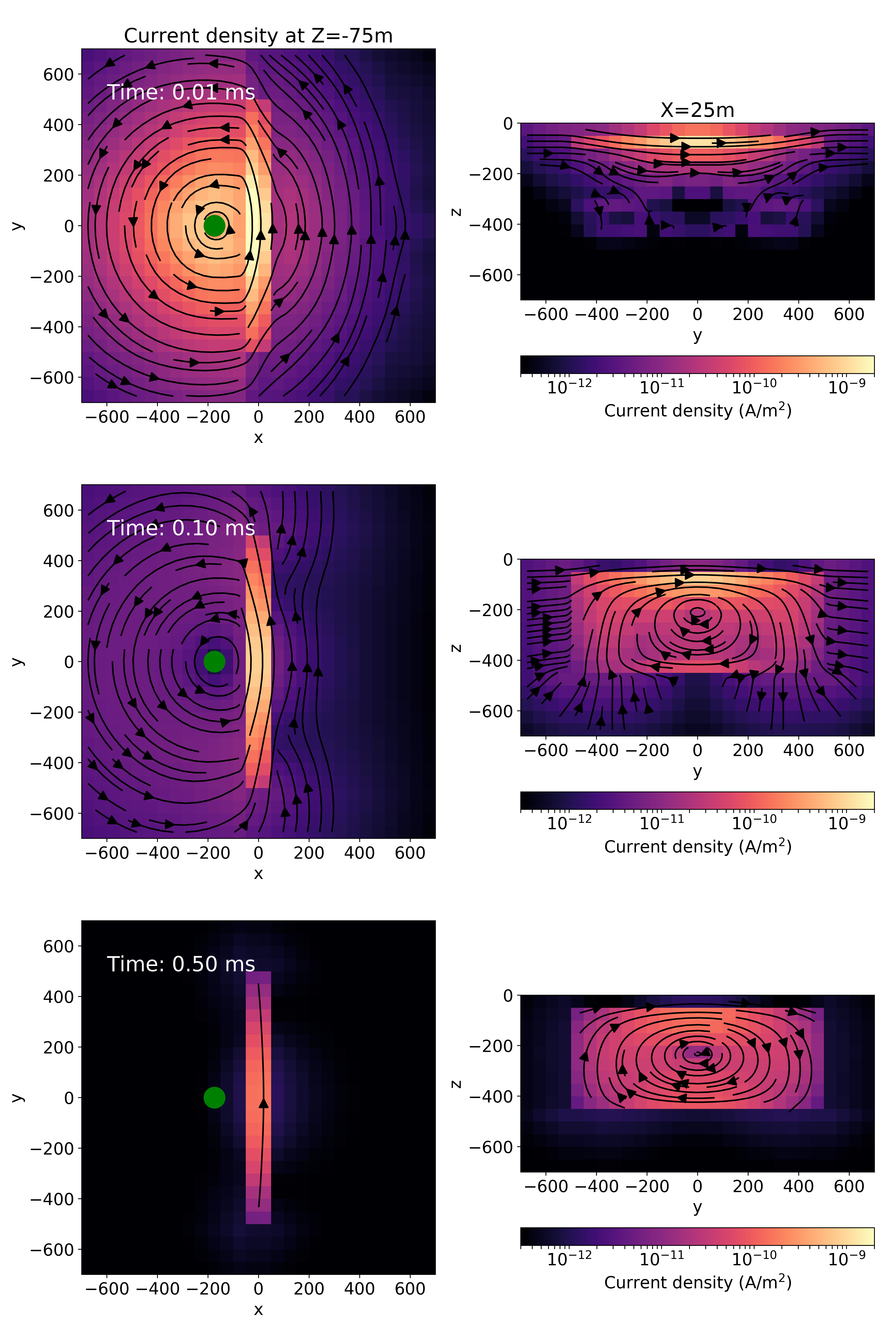}
  \caption{Depth slices (left) and cross-sections (right) showing the current density in the subsurface at 0.01 ms, 0.1ms and 0.5ms after shut-off. The source location is shown by the green dot.}
  \label{fig:currents}
\end{figure}

\begin{figure}[!htb]
  \centering
  \includegraphics[width=0.95\textwidth]{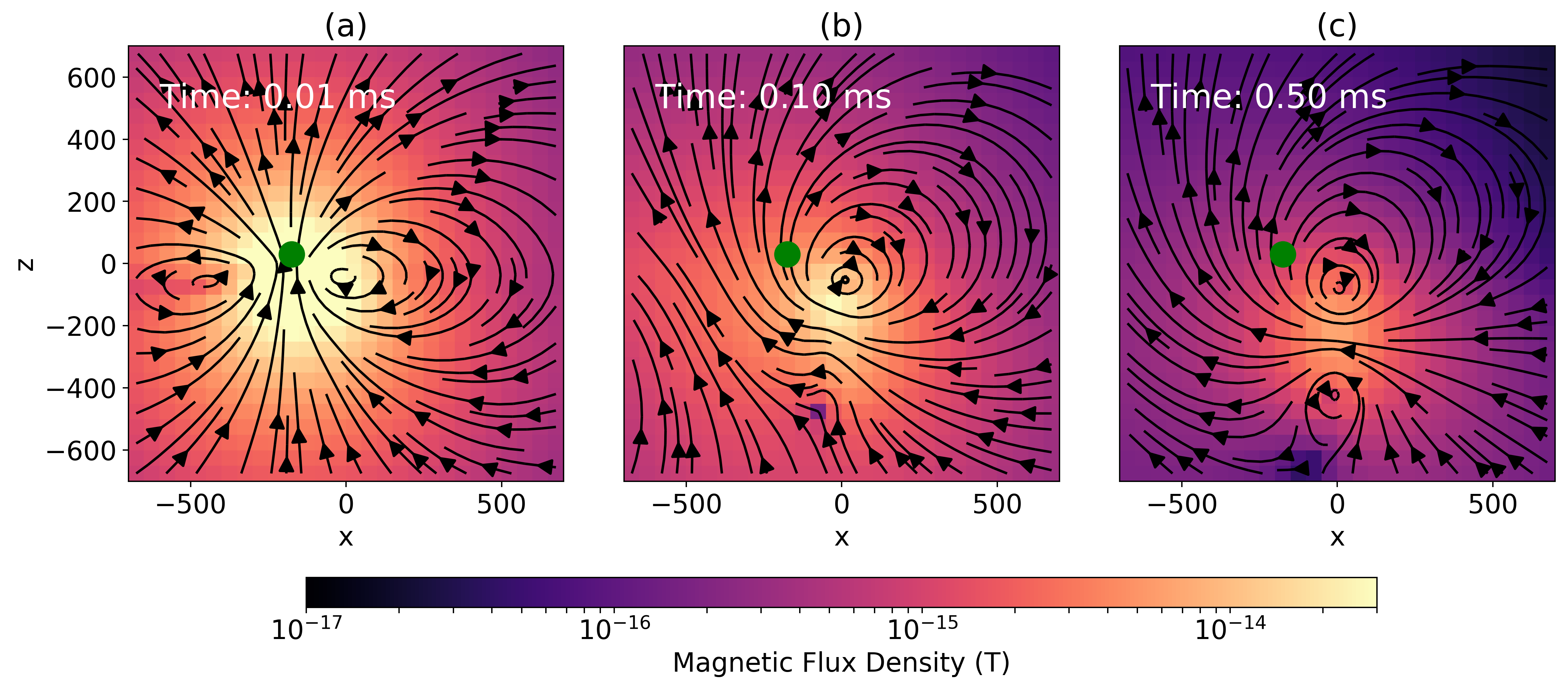}
  \caption{Magnetic flux density along the y=0m cross-section. The source and receiver location is shown by the green dot.}
  \label{fig:magnetic_flux}
\end{figure}

The synthetic data generated in this simulation are shown in Figure \ref{fig:data}; each profile line indicates $db_z/dt$ at a given time channel. At early times, we see a single peak anomaly over the plate and, as time progresses, a double-peak character emerges. To understand the responses, we plot the currents in the subsurface through time in Figure \ref{fig:currents} and the magnetic flux density in Figure \ref{fig:magnetic_flux}. The transmitter location is shown by the green dot.

Immediately after the source-current has been shut off, smoke-ring currents are induced in the earth. These horizontal, circular currents produce a magnetic field similar to that of a vertical magnetic dipole, as seen in Figure \ref{fig:magnetic_flux}a. When we are directly over the plate, which has a 100m width, the higher conductivity of the plate results in larger currents and thus a larger magnetic field. This is what gives us the single peak at early time in Figure \ref{fig:data}. The conductivity and thickness of the plate will control the time constant of the decay of these horizontal, circular currents.

At larger offsets, the plate still contributes to a larger signal as the smoke ring currents are channeled into the plate, as can be seen at t=0.01ms and t=0.1ms in Figure \ref{fig:currents}. The excessive currents are more like an electric dipole along the top of the plate, and this gives a magnetic field that is also pointing up at the receiver location; the resultant magnetic field can be seen in Figure \ref{fig:magnetic_flux}b, where the rotational magnetic fields at the top of the plate are generated by the channeled currents.

At intermediate times (0.1ms) the background field has set up strong vortex currents within the plate. By the time that the vortex currents are fully engaged, the galvanic currents are decaying, so at later times (0.5ms) the vortex currents dominate. The vortex currents effectively generate a horizontal magnetic dipole which has a positive vertical field at the receiver, as can be seen in Figure \ref{fig:magnetic_flux}c. It is this horizontal dipole contribution that leads to the double peak observed in the late-time data in Figure \ref{fig:data}.

\subsection{Stitched 1D inversion of AEM data}
When inverting airborne EM data, 1D inversions are commonplace. They have the benefit that each sounding is a relatively inexpensive computation and that the inversion can be readily parallelized over soundings. A basic 1D inversion will treat each sounding independently; more advanced 1D inversions may include lateral or spatial constraints in the regularization to promote smooth variation in the model between stations \citep{Viezzoli2008,Viezzoli2009}. For this example, we adopt the basic approach, inverting and regularizing each sounding independently. The extension to laterally or spatially constrained regularization can be considered within the SimPEG framework.

For each of the 1D inversions, the forward simulation is conducted on a cylindrically symmetric mesh. We perform a Tikhonov-style inversion \citep{Tikhonov1977}, which states the inverse problem as an optimization problem in which we minimize an objective function composed of a data-misfit term, $\phi_d$, and a regularization functional, $\phi_m$.
\begin{equation}
\begin{split}
&\underset{\mathbf{m}}{\text{minimize}} \quad \phi(\mathbf{m}) = \phi_d(\mathbf{m}) + \beta \phi_m(\mathbf{m}) \\
&\text{such that} \quad \phi_d \leq \phi_d^*
\end{split}
\label{eq:inverse_problem}
\end{equation}
The parameter $\mathbf{m}$ denotes the inversion model, which is the set of parameters sought in the inversion; $\beta$ is a trade-off parameter which weights the relative contribution of the data misfit and the regularization in the inversion. The mathematical description of the data misfit and regularization follow a standard $\ell_2$ formulation and are described in \cite{cockett2015}.

For the 1D inversions, the data were assigned 5$\%$ uncertainties and no noise floor was used. The trade-off parameter$\beta$, was set at a fixed value of 20. The starting and reference model were set to be equal to the background conductivity of $10^{-3}$ S/m. The inversions were terminated either when the target misfit was reached or 10 iterations were completed.

\begin{figure}[!htb]
  \centering
  \includegraphics[width=0.7\textwidth]{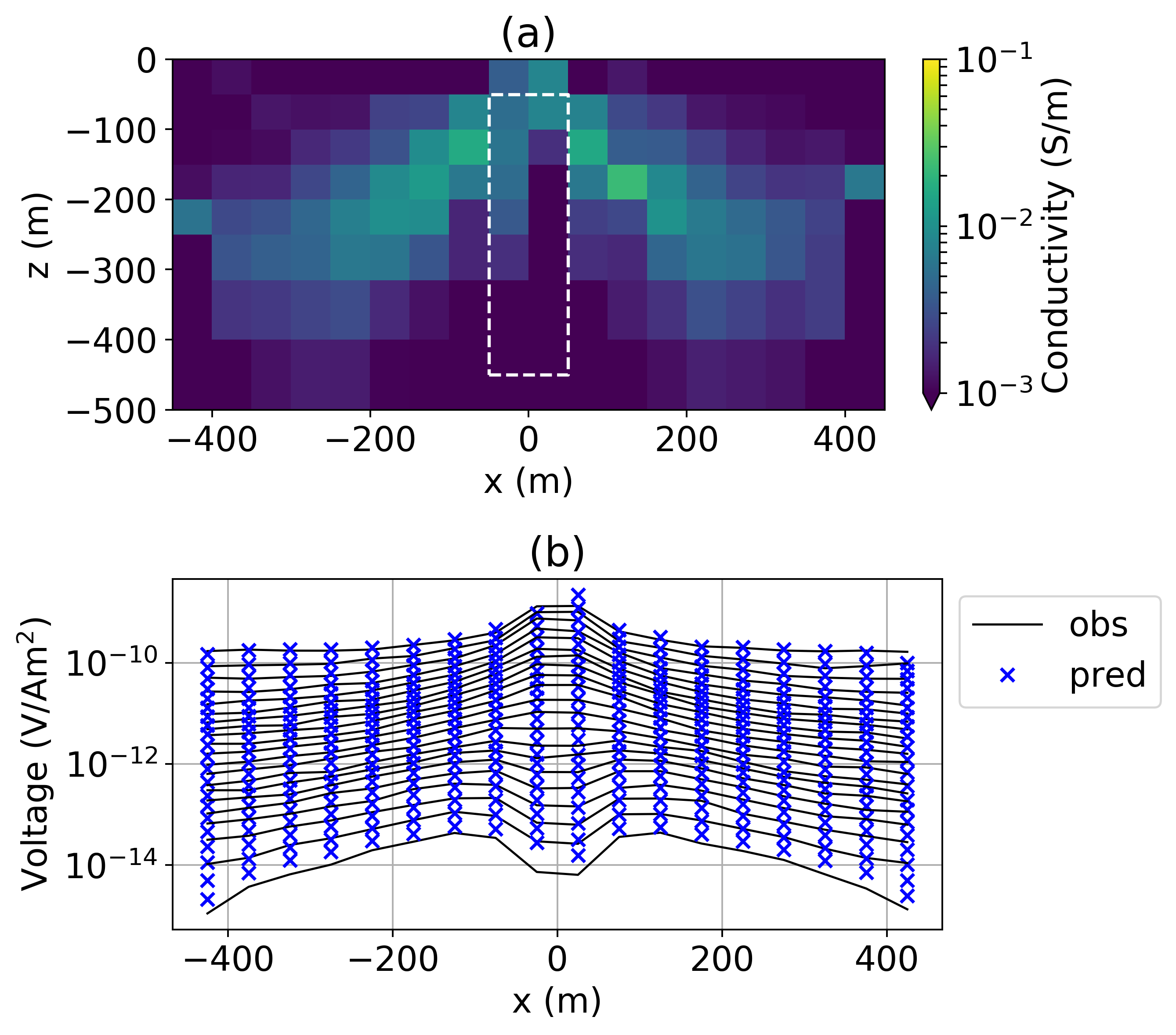}
  \caption{(top) Inverted 1D conductivity. (bottom) Predicted and observed data.}
  \label{fig:1Dinversion}
\end{figure}

The recovered model and associated predicted data are shown in Figure \ref{fig:1Dinversion}. Most of the soundings terminated at the maximum number of iterations without reaching the target misfit. The recovered conductivity model is typical of 2D effects in a 1D inversion: a resistor is imaged beneath conductive ``pant-legs'' which are centered above the vertical conductor. The 1D assumption is inadequate for imaging the 3D structure, as the horizontal-dipole behaviour observed in Figure \ref{fig:magnetic_flux}c cannot be explained by a 1D conductivity structure. In the next section, we will increase the dimensionality of the inversion and examine two approaches for a 2D inversion of these data.

\subsection{2D inversions of AEM data}

In both of the following inversions, the forward modelling is conducted on a 3D tensor mesh, while the inversion model we consider is 2D. To populate the 3D space with physical properties, the 2D slice is simply surjected along the y-axis; within SimPEG, this is handled by the Maps class. \cite{kang2015c} contains further discussion on the Maps class and how it appropriately modifies the sensitivity.

\begin{figure}[!htb]
  \centering
  \includegraphics[width=0.7\textwidth]{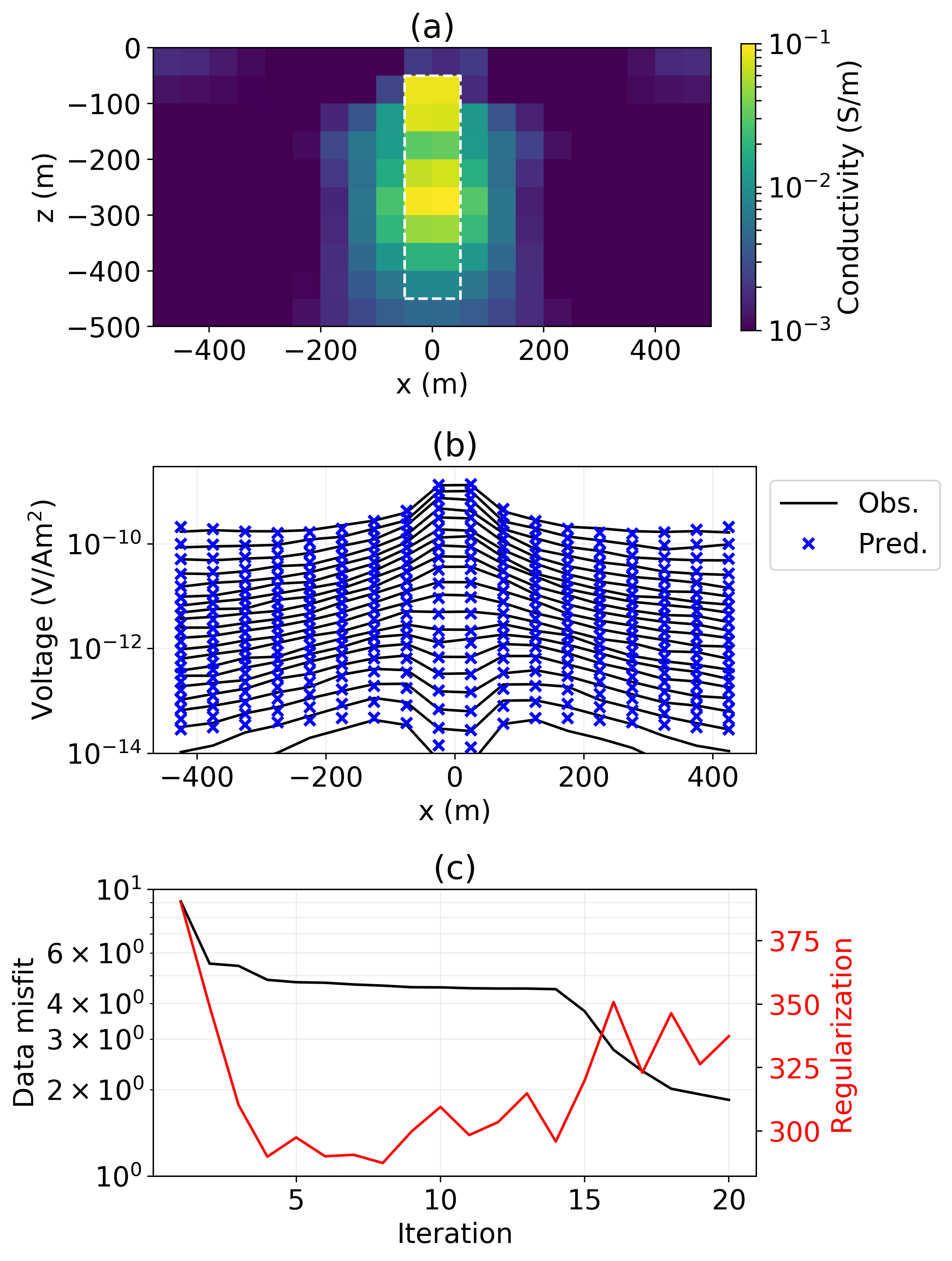}
  \caption{(a) Recovered conductivity model from the 2D TDEM inversion, (b) observed and predicted data, (c) data misfit and regularization over the course of the inversion.}
  \label{fig:2Dvoxel_inversion}
\end{figure}

The first approach we take to invert the data is a conventional voxel-based inversion. The data are assigned 5$\%$ uncertainties and a noise floor of $10^{-14}$ V/Am$^2$. A Tikhonov regularization is used and a beta-cooling schedule, which reduces the value of the trade-off parameter, $\beta$, by a factor of 5 every 3 iterations is employed. The starting model is a halfspace of $10^{-3}$ S/m. The inversion is run for 20 iterations and reaches an RMS misfit of 1.8. This is a reasonably good misfit considering we are fitting a 3D plate with a 2D model. The observed and predicted data are shown in Figure \ref{fig:2Dvoxel_inversion}b. We note how much better the two-peak portion of the data are fit compared to that obtained in the 1D. The recovered model is shown in Figure \ref{fig:2Dvoxel_inversion}. As is typical of smooth inversions, we recover a diffuse, conductive feature. This arises because we are penalizing gradients and using a fairly coarse mesh. By changing the norm used in the regularization, which can easily be done in SimPEG, sharper, more localized models can be obtained. Notwithstanding that, we note that the maximum amplitude of the recovered conductivity model closely reflects the true amplitude of the plate, and the horizontal location of the plate agrees well with the true model.

\begin{figure}[!htb]
  \centering
  \includegraphics[width=0.7\textwidth]{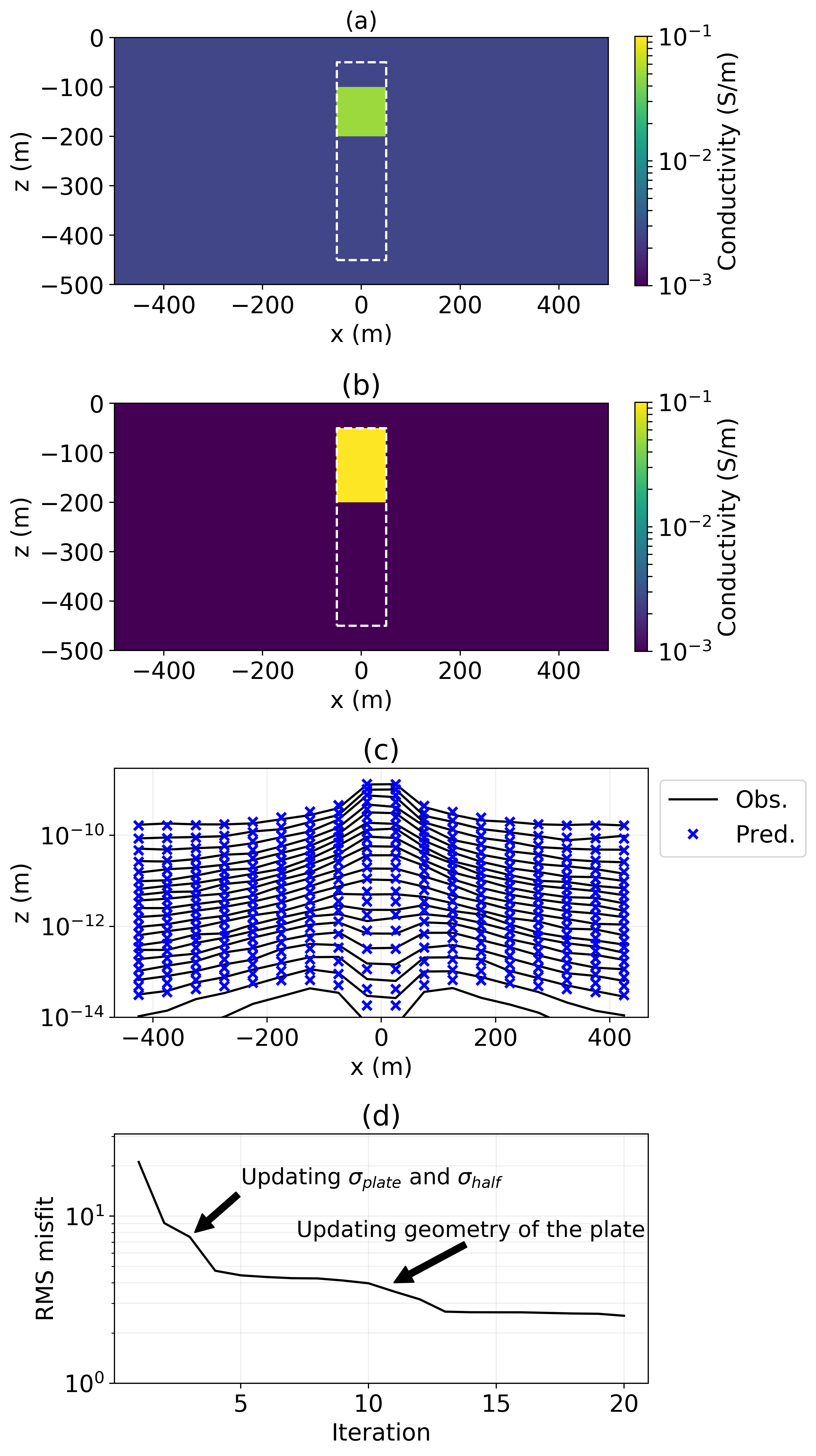}
  \caption{(a) Starting model for the parametric inversion, (b) conductivity model recovered in a parametric inversion for a 2D plate, (c) resultant predicted and observed data, (d) data misfit over the course of the inversion.}
  \label{fig:2Dparametric_inversion}
\end{figure}

An alternative approach to the inverse problem is to consider a parametric inversion. There are a number of motivations for doing this. First, we see from the data-misfit curve in Figure \ref{fig:2Dvoxel_inversion}c that this has been a challenging problem for the voxel inversion; between iterations 5 and 14, little progress is being made in reducing the misfit. This may be related to some of the challenges observed by \cite{mcmillan2015}, where the circular nature of the sensitivity of an airborne EM survey tends  to promote ``ringing'' artefacts in the inversion. The authors overcame this problem by introducing a parametric inversion. Another motivating factor for working in a parametric domain is that it may conform with \emph{a priori} knowledge about the earth.

For the above reasons we now consider a parametric approach to the inverse problem. This too, can be straightforwardly implemented in SimPEG. We describe the inversion model with 6 parameters: the conductivity of the background, the conductivity of the plate, the x and z locations of the center of the plate and its width in the x and z dimensions. The same data uncertainties as in the previous inversion are applied. No regularization is included, and the starting model used for the inversion is shown in Figure \ref{fig:2Dparametric_inversion}, the background conductivity is $5\times10^{-3}$S/m and the conductivity of the plate is 0.05 S/m. It is important to seed the parametric inversion with different background and plate conductivities; if the same conductivity is used for both, there is ambiguity introduced as the geometric parameters have no influence on the observed data if there is no conductivity contrast. The inversion is run for 20 iterations and the resultant model and data fit are shown in Figure \ref{fig:2Dparametric_inversion}. The RMS misfit of this inversion is 2.5.

The inversion accurately recovers the conductivities of the background and plate, the width of the plate and the location of the top of the plate. The depth extent is not resolved; we lose sensitivity with depth, and again the late-time data are poorly fit. This indicates that more progress might be made if we consider a 3D inversion. It is also interesting to examine the nature of the data misfit as a function of iteration and explore which parameters are updated at each iteration. In the first few iterations, the inversion updates the conductivity of the background and the plate. Once it has made sufficient progress on these parameters, it then proceeds to update the geometric properties. The order-and-magnitude of the updates at each step provide some indication of the sensitivity of the inversion to each parameter; this can be useful in a feasibility study to assess how well we might expect the inversion to resolve a geologic feature of interest.
A similar RMS misfit is reached in both inversions, and the data are fit quite well, with the exception of the last two time channels. This is likely due to the 3D nature of the target. Further improvements to the data fit could be explored with 3D voxel or parametric inversions within the SimPEG framework.

\section{Outlook}
Using an example of a conductive, vertical plate in a resistive half-space, we demonstrated the use of SimPEG for: (a) building up an understanding of the observed data by visualizing the behaviour of electromagnetic fields and fluxes and (b) for exploring approaches for the challenging inverse problem of inverting for a compact, conductive target in airborne EM data. The scripts used to generate the figures shown in this abstract are freely available so that they can be reproduced and further explored. In developing SimPEG as an open source project, we aim to create opportunities for collaboration and the integration of expertise from a diverse group of researchers. As we move forward with the electromagnetics module, we plan to scale to larger 3D inversion by implementing domain decomposition similar to \cite{yang2014}. Due to the modular nature of SimPEG, this extension will be applicable for tensor, OcTree and cylindrical meshes. We welcome new contributions and hope that these efforts promote more reproducibility and transparency in geophysical simulations and inversions.

\section{Acknowledgements}
The authors thank the SimPEG community including Dom Fournier, Gudni Rosenkjaer, Thibaut Astic, Michael Mitchell, Joseph Capriotti, and David Marchant for their contributions which have improved the quality of the software and extended its capabilities. Thanks also to Dr. Eldad Haber for his guidance and instruction on finite volume techniques and the implementation of the electromagnetic problem. We are grateful to Dr. Niels Christensen and the two anonymous reviewers for their constructive feedback on the manuscript.

%% ==================================================================
\clearpage
\bibliographystyle{seg}  % style file is seg.bst
\bibliography{biblio}
\end{document}